\documentclass{article}
\usepackage{dahlin,testJD,enumitem,hyperref}           
\usepackage{bm}

\title{Particle filter-based Gaussian process optimisation for parameter inference}
\author{Johan Dahlin and Fredrik Lindsten%
\thanks{Supported by the project Probabilistic modelling of dynamical systems (Contract number: 621-2013-5524) funded by the Swedish Research Council. JD is with the Division of Automatic Control, Link{\"o}ping University, Link{\"o}ping, Sweden. E-mail: \texttt{  johan.dahlin@liu.se}. FL is with the Department of Engineering, University of Cambridge, Cambridge, United Kingdom. E-mail: \texttt{fredrik.lindsten@eng.cam.ac.uk}}%
}

\begin{document}
\maketitle
\thispagestyle{empty}
\pagestyle{empty}

\begin{abstract}
We propose a novel method for maximum likelihood-based parameter inference in nonlinear and/or non-Gaussian state space models. The method is an iterative procedure with three steps. At each iteration a particle filter is used to estimate the value of the log-likelihood function at the current parameter iterate. Using these log-likelihood estimates, a surrogate objective function is created by utilizing a Gaussian process model. Finally, we use a heuristic procedure to obtain a revised parameter iterate, providing an automatic trade-off between exploration and exploitation of the surrogate model. The method is profiled on two state space models with good performance both considering accuracy and computational cost.
\end{abstract}

\section{Introduction}
We are interested in maximum likelihood-based (ML) parameter inference in nonlinear and/or non-Gaussian state space models (SSM). An SSM with latent states $\statesdef$ and measurements $\measurementsdef$is defined as
\begin{subequations}
\begin{align}
	x_{t}|x_{t-1}  &\sim  f_{\theta}(x_{t}|x_{t-1}), \\
	y_{t}|x_t      &\sim  g_{\theta}(y_{t}|x_t),
\end{align}%
\label{eq:SSM}%
\end{subequations}%
\noindent where $f_{\theta}(\cdot)$ and $g_{\theta}(\cdot)$ denote known distributions parame- trised by the unknown static parameter vector $\theta \in \Theta \subseteq \mathbb{R}^d$. For simplicity, we assume that the initial state $x_0$ is known. Let $\Lfunc[] \triangleq p_\theta(\measurements)$ denote the likelihood of $\measurements$ for a given value of $\theta$. In ML estimation, we wish to estimate $\theta$ by solving the optimisation problem,
\begin{align}
	\mletheta = \argmax_{\theta \in \Theta } \Lfunc[] = \argmax_{\theta \in \Theta } \ell(\theta),
	\label{eq:MLEoptimisation}
\end{align}
where $\ell(\theta) \triangleq \log \Lfunc[]$ denotes the log-likelihood function. Extensive treatments on ML inference are found in e.g.\ \cite{Ljung1999} and \cite{LehmannCasella1998}.

The likelihood for a general SSM can be expressed as
\begin{align}
	\Lfunc[]
	= 
	\prod_{t=1}^T
	p_{\theta}(y_t|y_{1:t-1}),
	\label{eq:likeFunc}
\end{align}
where $p_{\theta}(y_t|y_{1:t-1})$ denotes the one-step predictive density. For a linear Gaussian models, these densities can be computed exactly by using the Kalman filter. However, for a nonlinear model the one-step predictive densities are in general intractable. It is therefore also intractable to evaluate the objective function in \eqref{eq:MLEoptimisation}, which poses an obvious difficulty in addressing the ML problem.

Recently, ML estimation has been carried out in nonlinear SSMs by the aid of Sequential Monte Carlo \cite{DoucetJohansen2011}. This includes e.g.\ using gradient-based search \cite{PoyiadjisDoucetSingh2011} and the Expectation Maximisation (EM) algorithm \cite{SchonWillsNinness2011,Lindsten2013}. However, some of these methods require computationally costly particle smoothing to estimate the necessary quantities, which can be a problem in some situations.

An alternative is to make use of the simultaneous perturbation stochastic approximation (SPSA) algorithm \cite{Spall1987}, which uses a steepest ascent algorithm with a stochastic approximation scheme to estimate the solution to \eqref{eq:MLEoptimisation}. The gradients are estimated using finite differences with random perturbations. This results in that the algorithm only needs to sample the likelihood function twice at each iteration, independent of the dimension of the problem. SPSA is used in combination with SMC in e.g.\ \cite{SinghWhiteleyGodsill2013} and \cite{EhrlichJasraKantas2012}. 

Another approach for maximum likelihood estimation is based on approximate inference based on Laplace approximations and moment matching. We do not consider these methods any further in this paper and refer interested readers to e.g.\ \cite{Bishop2006}, \cite{KhanMohamedMurphy2012} and \cite{Bell2000} for more information.

In this paper, we propose a novel algorithm for ML estimation of static parameters in a nonlinear SSM. The method combines particle filtering (PF) with Gaussian process optimisation (GPO) \cite{Jones2001,Boyle2007,Lizotte2008}. The latter is a method well-suited for optimisation when it is costly to evaluate the objective function. The resulting algorithm is efficient in the sense that it provides accurate parameter estimates while making use of only a small number of (costly) log-likelihood evaluations.

\section{Maximum likelihood estimation with a surrogate cost function}
\label{sec:overview}
We now turn to our new procedure for ML estimation of general nonlinear SSMs \eqref{eq:SSM}. We start by outlining the main ideas of the procedure on a high level. The individual steps of the algorithm are discussed in detail in the consecutive sections. The algorithm is an iterative procedure, which thus generates a sequence of iterates $\{\theta_k \}_{k\geq 0}$ for the model parameters. Each iteration consists of three main steps:

\begin{enumerate}
\item[(i)] Given the current iterate $\theta_k$, compute an estimate of the objective function (i.e.\ the log-likelihood) for this parameter value, denoted as $\widehat\ell_k \approx \ell(\theta_k)$.
\item[(ii)] Given the collection of tuples $\{ \theta_j, \widehat\ell_j \}_{j=0}^k$ generated up to the current iterate, create a
  model of the (intractable) objective function $\ell(\theta)$.
\item[(iii)] Use the model as a surrogate for the objective function to generate a new iterate $\theta_{k+1}$.
\end{enumerate}
Note that the method requires only one estimation of the log-likelihood function at each iteration. This is promising, since it is typically computationally costly to estimate the log-likelihood value and we therefore wish to keep the number of such evaluations as low as possible.

For step~(i), i.e.\ evaluating the log-likelihood function for a given value of $\theta$, we use a PF, resulting in a (noisy) estimate of the objective function. This step is discussed in Section~\ref{sec:smc}. For steps~(ii) and~(iii), we apply the GPO framework. First, we construct a surrogate for the objective function by modelling it as a Gaussian process, taking the information available in the previous iterates $\{ \theta_j, \widehat\ell_j \}_{j=0}^k$ into account. This is discussed in Section~\ref{sec:gp}. 

Then, we make use of a heuristic, referred to as an \emph{acquisition rule}, to find the next iterate $\theta_{k+1}$ based on the GP model. The acquisition rule is such that it favours values of $\theta$ for which the model predicts a large value of the objective function and/or where there is a high uncertainty in the model. This is useful since it automatically results in a trade-off between exploration and exploitation of the model.

In this paper, we consider a simple numerical example to illustrate the different steps of the algorithm during the derivation. For this, the linear Gaussian state space (LGSS) model,
\begin{subequations}
\begin{align}
	x_{t+1}|x_t  &\sim  \mathcal{N} \left( x_{t+1}; \theta x_t, 1 \right), \\
	y_{t}|x_t    &\sim  \mathcal{N} \left( y_t; x_t, 0.1^2 \right),
\end{align}%
\label{eq:LGSSMmodel}%
\end{subequations}%
\noindent with $\Theta = [-1,1]$ and parameter $\theta^{\star}=0.5$ is simulated for $T=250$ time steps. The complete algorithm is evaluated in Section \ref{sec:results} on this model, as well as on a nonlinear SSM.

\section{Estimating the log-likelihood}
\label{sec:smc}

We begin this section with a brief description of a PF. For more general introductions, see e.g. \cite{DoucetJohansen2011}. We then continue with discussing the specific problem of likelihood estimation using the PF.

\subsection{The particle filter}
The PF is a sequential Monte Carlo method used to approximate e.g.\ the intractable filtering distribution $\mfilterdist$ for a general SSM \eqref{eq:SSM}. This is done by representing it by a set of $N$ weighted particles $\partsys$ according to
\begin{align*}
	\mfilterdistpart 
	\triangleq 
	\sum_{i=1}^N 
	\frac 
	{ \fweighti  }
	{ \sum_{k=1}^N w_{t}^{(k)} }
	\delta_{\parti} (\dn x_t),
\end{align*}
where $\fweighti$ and $\parti$ denote the (unnormalised) weight and state of particle $i$ at time $t$, respectively. Here, $\delta_z(\dn x_t)$ denotes the Dirac measure located at the point $z$. These approximations are generated sequentially in time $t$. Given the particles at time $t-1$, the PF proceeds to time $t$ by: (a) resampling, (b) propagation and (c) weighting.

In step (a), the particles are resampled with replacement, using the probabilities given by their (normalized) importance weights. This is done to rejuvenate the particle system and to put emphasis on the most probable particles. The result is an unweighted particle system $\{\widetilde{x}_{t-1}^{(i)}, 1/N \}_{i=1}^N$, targeting the same distribution $p_{\theta}(x_{t-1}|y_{1:t-1})$.

In step (b), the particles are propagated to time $t$ by sampling from a proposal kernel $\parti \sim R_{\theta} \big( x_t | \widetilde{x}_{t-1}^{(i)}, y_t \big)$ from $i = 1$ to $N$. Finally in Step (c), the particles are assigned importance weights. This is done to account for the discrepancy between the proposal and the target densities. The importance weights are given by
\begin{align}
    w_{t}^{(i)} 
	= 
	W_{\theta}(x_t^{(i)}, \widetilde{x}_{t-1}^{(i)})
	=
	\frac
	{ g_{\theta}(y_t|x^{(i)}_t) f_{\theta}(x^{(i)}_t|\widetilde x_{t-1}^{(i)}) }
	{ R_{\theta} \Big (x^{(i)}_t|\widetilde x_{t-1}^{(i)}, y_t \Big) }.
	\label{eq:APFweights}
\end{align}
In the sequel, we use the \textit{bootstrap} PF which means that new particles are proposed according to the state dynamics, i.e.\ $R_{\theta}(\cdot) = f_{\theta}(\cdot)$ and $w^{(i)}_{t} = g_{\theta}(y_t|x_t^{(i)})$. Although more sophisticated alternatives exist, see e.g.\ the fully-adapted PF introduced in \cite{PittShephard1999}.

\subsection{Estimation of the likelihood}
In order to use the PF for estimating the likelihood, we start by writing the one-step predictive density as
\begin{align*}
	&p_{\theta}(y_{t}|y_{1:t-1}) 
	= 
	\dint
	p_{\theta}(y_t,x_t|x_{t-1})
	p_{\theta}(x_{t-1}|y_{1:t-1}) 
	\dd x_{t-1:t} \nonumber \\
	&=
	\dint
	W_{\theta}(x_t, x_{t-1})  R_{\theta}(x_t|x_{t-1},y_t)
	p_{\theta}( x_{t-1} |y_{1:t-1}) 
	\dd x_{t-1:t},
\end{align*}
where we have multiplied and divided with the proposal kernel $R_{\theta}(\cdot)$. To approximate the integral, we note that the (unweighted) particle pairs $\{ \widetilde x_{t-1}^{(i)}, x_t^{(i)} \}_{i=1}^N$ are approximately drawn from $R_{\theta}(x_t|x_{t-1},y_t) p_{\theta}( x_{t-1} |y_{1:t-1}) $. Consequently, we obtain the Monte Carlo approximation 
\begin{align*}
p_{\theta}(y_{t}|y_{1:t-1}) \approx \frac{1}{N} \sum_{i=1}^N  w^{(i)}_{t}.
\end{align*}
By inserting this approximation into \eqref{eq:likeFunc} we obtain the particle estimate of the likelihood,
\begin{align*}
	\LfuncEst[]
	=
	\prod_{t=1}^T \left(  \frac{1}{N} \sum_{i=1}^N w_{t}^{(i)} \right).
\end{align*}
This likelihood estimator has been studied extensively in the SMC literature. The estimator is consistent and, in fact, also unbiased for any $N \geq 1$; see e.g.\ \cite{PittSilvaGiordaniKohn2012} and Proposition 7.4.1 in \cite{DelMoral2004}. Furthermore, a central limit theorem holds,
\begin{align*}
  \sqrt{N} \left[ \LfuncEst[] - \Lfunc[] \right] 
  \stackrel{d}{\longrightarrow} 
  \mathcal{N} \Big( 0,\psi^2(\theta) \Big),
\end{align*}
for some asymptotic variance $\psi^2(\theta)$; see Proposition 9.4.1 in \cite{DelMoral2004}.

\begin{algorithm}[!t]
\caption{\textsf{PF for log-likelihood estimation}}
\small
\textsc{Inputs:} An SSM \eqref{eq:SSM}, $\measurements$ (obs.) and $N$ (no.\ particles). \\
\textsc{Output:} $\LLfuncEst[]$ (est.\ of the log-likelihood). 
\algrule[.4pt]
\begin{algorithmic}[1]
	\STATE Initialise particles $x_0^{(i)}$ for $i=1$ to $N$. 
	\FOR{$t=1$ to $T$}
		\STATE Resample the particles with weights $\{w_{t-1}^{(i)}\}_{i=1}^N$.
		\STATE Propagate the particles using $R_{\theta}(\cdot)$.
		\STATE Compute \eqref{eq:APFweights} to obtain $\{w_{t}^{(i)}\}_{i=1}^N$.
	\ENDFOR
	\STATE Compute \eqref{eq:LLikeEst} to obtain $\LLfuncEst[]$.
\end{algorithmic}
\label{alg:APFlike}
\end{algorithm}

\subsection{Estimation of the log-likelihood}
However, working directly with the likelihood typically results in numerical difficulties. To avoid problems with numerical precision, we instead use an estimate of the log-likelihood
\begin{align}
	\LLfuncEst[]
	= 
	\log \widehat{\mathcal{L}}(\theta)
	=
	\sum_{t=1}^T \log \left[ \sum_{i=1}^N w_{t}^{(i)} \right] - T \log N.
	\label{eq:LLikeEst}
\end{align}
The resulting complete algorithm for estimating the log-likelihood using a PF is presented in Algorithm \ref{alg:APFlike}.

Note that, by taking the logarithm of $\LfuncEst[]$, we introduce a bias into the estimator. However, by the second-order delta method \cite{CasellaBerger2001}, the asymptotic normality carries over to the log-likelihood estimate,
\begin{align}
  \sqrt{N} \left[ \LLfuncEst[] - \LLfunc[] \right] 
  \stackrel{d}{\longrightarrow} 
  \mathcal{N} \left( 0 ,\gamma^2(\theta) \right),
  \label{eq:llerrordist}
\end{align}
where $\gamma(\theta) = \psi(\theta)/\mathcal{L}(\theta)$. Motivated by this, we make the assumption that the log-likelihood estimates are Gaussian distributed and centered around the true log-likelihood value.
That is, we can write
\begin{align}
	\LLfuncEst[] = \ell(\theta) + z, \qquad z \sim \mathcal{N}(0,\sigma^2_z).
        \label{eq:GaussAssumption}
\end{align}
Similar normality assumptions have previously been used by \cite{PittSilvaGiordaniKohn2012} and \cite{DoucetPittKohn2012}.
The unknown variance $\sigma^2_z$ is treated as a free parameter that is estimated on-the-fly as we run the proposed estimation algorithm. That is, we do \emph{not} have to estimate $\sigma_z^2$ by making any initial test runs. We return to this in the sequel.

We validate the Gaussian assumption \eqref{eq:GaussAssumption} using a small numerical experiment to illustrate the bias and variance, at a finite number of particles. We calculate $1 \thinspace 000$ estimates of the log-likelihood $\ell(0.5)$ for the model in \eqref{eq:LGSSMmodel}. This is done by running Algorithm~\ref{alg:APFlike} independently $1\thinspace000$ times with $N = 1\thinspace000$
particles.

\begin{figure}[ht]
	\centering
	\includegraphics[width=\columnwidth]{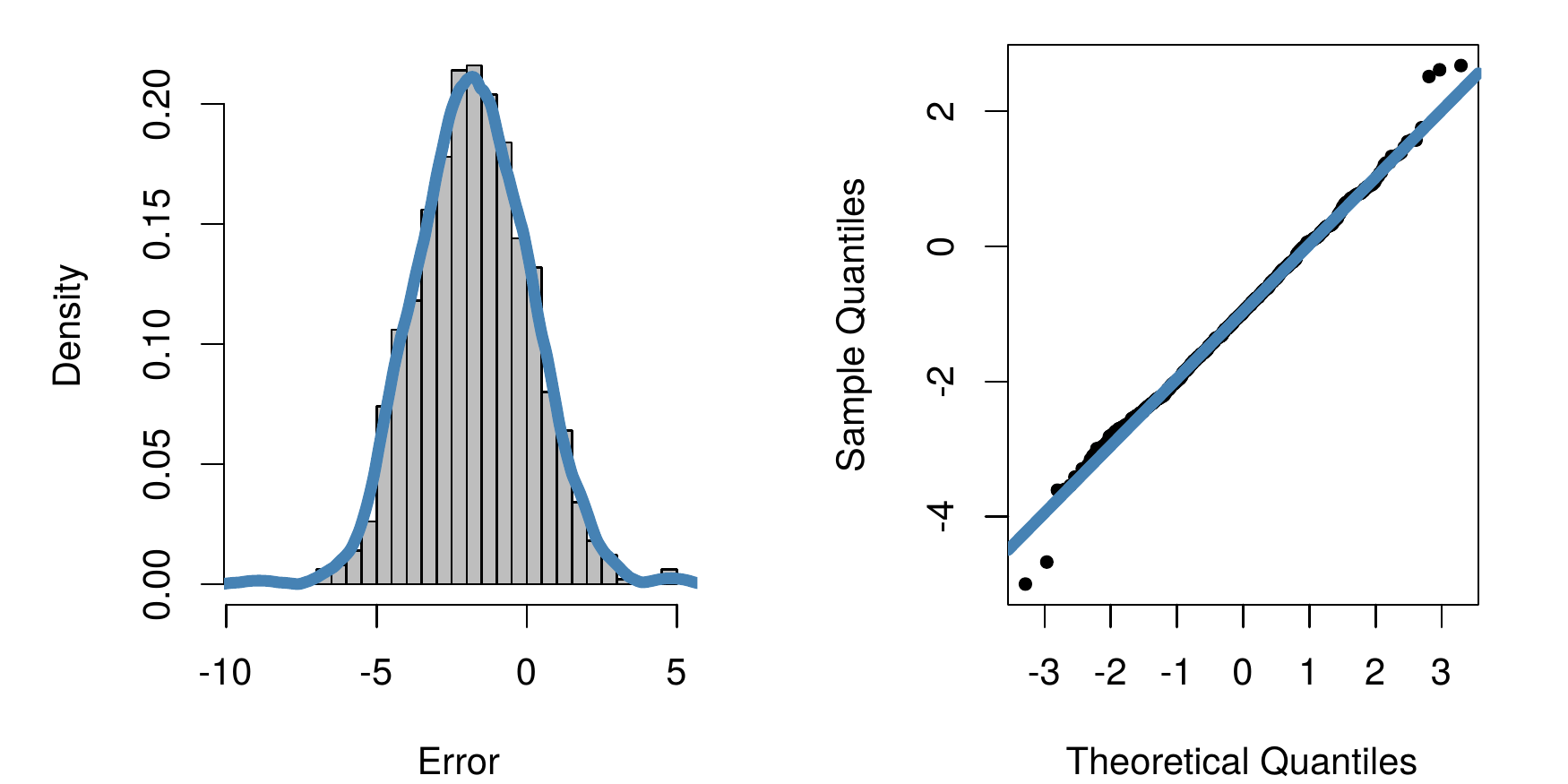} 
	\caption{\small{Left: the histogram and kernel density estimate (blue line) of the estimation error of the log-likelihood in the LGSS model \eqref{eq:LGSSMmodel} at $\theta=\theta^{\star}$. Right: the QQ-plot of the data with the theoretical quantiles marked with the solid blue line.}}
	\label{fig:LLestLGSSM}
\end{figure}

In Figure \ref{fig:LLestLGSSM}, we present the distribution of the error in the estimates together with a QQ-plot. Both plots validate that the estimates are approximatively distributed according to a Gaussian distribution. Also a Lilliefors hypothesis test \cite{Lilliefors1967} does not reject the null hypothesis, that the measurements are drawn from a Gaussian distribution at significance level $\alpha=0.05$.

\section{Modelling the surrogate function}\label{sec:gp}
From the previous, we consider a naive approach to solve \eqref{eq:MLEoptimisation} by creating a grid of the parameter space and estimating the log-likelihood in each grid point. The parameter estimate is then obtained as the grid point that maximises the objective function. The problem here is that as the dimension of the parameter space increases, an exponentially increasing number of grid points is required to retain the accuracy of the estimate. 

Furthermore, using finite differences to compute the gradient of the log-likelihood is problematic due to the noise in \eqref{eq:GaussAssumption}. This problem can be mitigated by using a particle smoother, as previously discussed in e.g.\ \cite{PoyiadjisDoucetSingh2011}, but this is even more computationally expensive than running the particle filter. Instead, we construct a model of the noisy log-likelihood evaluations in Step (ii). This model then serves as a surrogate for the actual objective function.

\subsection{Gaussian process model}
In this paper, we use a GP for this purpose, as these processes are possibly flexible enough to capture the overall structure of the log-likelihood for many SSMs. GPs can be seen as a generalisation of the multivariate Gaussian distribution and are commonly used as \textit{priors over functions}. In this view, the resulting posterior obtained by conditioning upon some observations, describes the functions that could have generated the observations. This makes GPs a popular class of nonparameteric models used for e.g.\ regression, classification and optimisation, see e.g.\ \cite{RasmussenWilliams2006} and \cite{Murphy2012}.

In the following, we model the log-likelihood $\LLfunc[]$ as being \emph{a priori} distributed according to a GP. That is,
\begin{align}
	\ell(\cdot)
    \sim
    	\mathcal{GP} \Big( m(\cdot), \kf(\cdot,\cdot) \Big),
	\label{eq:GPdef}
\end{align}
where the process is fully described by the mean function $m(\cdot)$ and the covariance function $\kf(\cdot,\cdot)$.

\subsection{Updating the model and the hyperparameters}
To ease the presentation, we here consider a particular iteration $k$ of the GP and the PF. Let $\mathcal{D}_k=\{\bm{\theta}_k,\widehat{\bm{\ell}}_k\}=\{\theta_j,\widehat\ell(\theta_j)\}_{j=1}^k$ denote a set of iterates, where $\bm{\theta}_k$ and $\widehat{\bm{\ell}}_k$ denote vectors obtained by stacking the $k$ parameters and noisy log-likelihood estimates, respectively.

It follows that the \textit{posterior distribution} is given by
\begin{align}
		\LLfunc[] | \mathcal{D}_k
        \sim	
        \mathcal{N} \Big( \mu(\theta| \mathcal{D}_k), \sigma^2(\theta| \mathcal{D}_k) + \sigma^2_z \Big),
        	\label{eq:GPpost}
\end{align}
where $\mu(\theta| \mathcal{D}_k)$ and $\sigma^2(\theta| \mathcal{D}_k)$ denote the posterior mean and variance given the iterates $\mathcal{D}_k$, respectively. By standard results for the Gaussian distribution, we have
\begin{subequations}
\begin{align}
	\mu( \theta | \mathcal{D}_k )
	&=
	m(\theta)
	+
	\kf( \theta , \bm{\theta}_k )
	\Gamma^{-1}
	\left[
	\widehat{\bm{\ell}}_k
	- 
	m(\theta)
	\right], \\
	\sigma^2( \theta | \mathcal{D}_k )
	&=
	\kf( \theta , \theta )
	-
	\kf( \theta, \bm{\theta}_k )
	\Gamma^{-1}
	\kf( \bm{\theta}_k , \theta ),
\end{align}%
\label{eq:GPpred}%
\end{subequations}%
\noindent with $\Gamma = \kf( \bm{\theta}_k , \bm{\theta}_k ) + \sigma^2_z	\mathbf{I}_{k \times k}$, and where $\mathbf{I}_{k \times k}$ denotes a $k \times k$-identity matrix. Here we note that the posterior distribution can be sequentially updated to save computations, see the aforementioned references for details. 

In the GP model presented, we use some mean function and covariance function that possibly depend on some unknown hyperparameters. Also, we need to estimate the unknown noise variance $\sigma^2_z$ in \eqref{eq:GaussAssumption}. For this, we adopt the emperical Bayes (EB) procedure to estimate these quantities. This is done by numerically optimising the marginal likelihood of the data with respect to the hyperparameters.

\subsection{Example of log-likelihood modelling}
\begin{figure}[!t]
	\centering
	\includegraphics[width=\columnwidth]{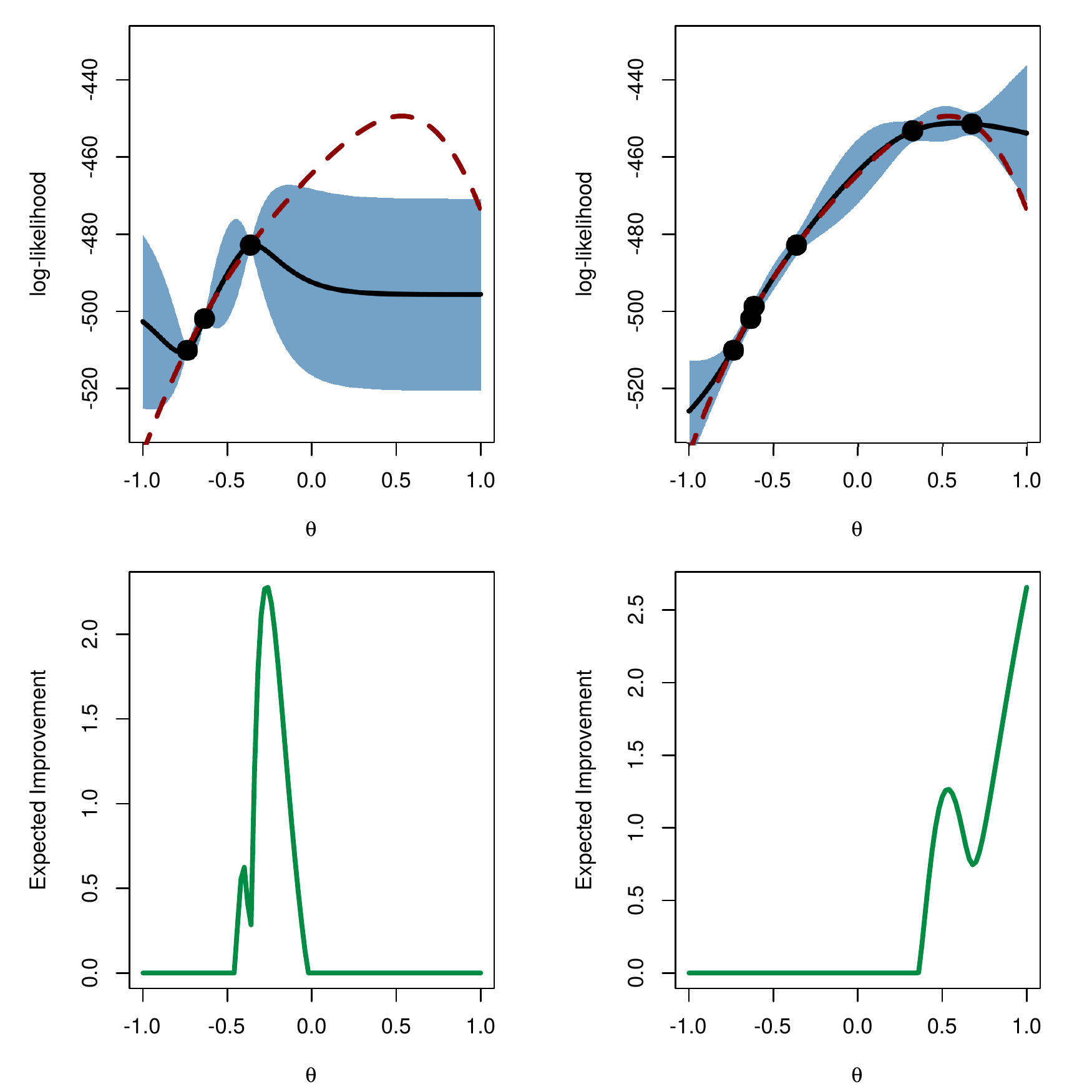} 
	\caption{\small{Upper: The surrogate function of the LGSS model \eqref{eq:LGSSMmodel} using three (left) and six (right) uniform samples "$\bullet$", respectively. The solid line presents the value of the predictive mean function with its $95\%$ CI in blue and the dashed red line presents the true likelihood. Lower: The corresponding EIs using $\zeta=0.01$.}}
	\label{fig:GPexample}
\end{figure}

We end this section by an example to illustrate the usefulness of GPs in modelling the log-likelihood. In the upper part of Figure \ref{fig:GPexample}, we show the posterior distribution of the log-likelihood of the model in \eqref{eq:LGSSMmodel}. The posterior is estimated using three (left) and six (right) samples of the log-likelihood drawn at some randomly selected parameters. With information from only six samples, the mean of the surrogate function passes close to the observed iterates with a reasonable confidence interval.

\section{Acquisition rules}
\label{sec:acquisitionrules}
The remaining problem in the proposed algorithm is how to select the parameters at which the log-likelihood should be evaluated in step (iii). A simple choice would be to consider a random sampling approach, which works well when the dimension of the parameters is small. However, when the dimension increases, we are faced with the curse-of-dimensionality and independent sampling is inefficient. 

As previously discussed, we instead use acquisition rules that balances exploration and exploitation of the parameter space and makes use of the posterior distribution obtained from the GP. These heuristics are well-studied in GPO and simulation-based comparisons are presented in e.g.\ \cite{Lizotte2008}. In this paper, we follow their general recommendations and use the expected improvement (EI) from \cite{Jones2001}.

\subsection{Expected improvement}
Consider the \textit{predicted improvement} defined as
\begin{align}
I(\theta) &= \max \Big\{0, \ell(\theta) - \mu_{\max} - \zeta \Big\},
\end{align}
where $\zeta$ is a user-defined coefficient that balances exploration and exploitation. Also, introduce the expected peak of the log-likelihood function,
\begin{subequations}
\begin{align}
\mu_{\max} &= \max_{\theta \in \bm{\theta}_k } \mu(\theta|\mathcal{D}_k),
\end{align}%
\label{eq:mumax}%
\end{subequations}%
\noindent over the previous iterates. Here, we again consider a particular iteration $k$ in the notation for brevity. 

Finally, by using the posterior distribution obtained from the GP, we can write the EI as
\begin{align}
    \mathbb{E}[I(\theta) | \mathcal{D}_k] 
	&= \sigma(\theta) \Big[ Z(\theta) \Phi \big( Z(\theta) \big) + \phi \big( Z(\theta) \big) \Big], \text{ with }
	\label{eq:EIquantity} \\
	Z(\theta) &=  \sigma^{-1}(\theta) \Big[ \mu(\theta) - \mu_{\max} - \zeta \Big], \nonumber
\end{align}
where we drop the dependence on $\mathcal{D}_k$ for brevity. Here, $\Phi$ and $\phi$ denote the CDF and PDF of the standard Gaussian distribution, respectively. An acquisition rule follows by the maximising argument
\begin{align}
	\theta_{k+1} = \argmax_{\theta\in \Theta} \mathbb{E} \Big[ I(\theta) | \mathcal{D}_k \Big] ,
	\label{eq:EIdef}
\end{align}
i.e.\ we sample the likelihood in $\theta_{k+1}$ during the next iteration of the algorithm. 

In the lower part of Figure \ref{fig:GPexample}, the expected improvements are shown for the situation discussed in the previous example. The two situations correspond to an exploitation step (left) and an exploration step (right), respectively. In the former, we sample in the neighbourhood of the current predicted peak. In the latter, we sample in an area where the uncertainty is large to determine if there is a peak in that area. 

From the expression in \eqref{eq:EIquantity}, we expect a high value of EI for parameters where the variance $\sigma(\theta)$ is large. If also the predictive mean $\mu(\theta)$ is larger than $\mu_{\max}$, then the EI assumes even larger values for these parameters. This gives the desired behaviour of the acquisition function discussed previously.

\begin{algorithm}[!t]
\caption{\textsf{Particle-based parameter inference in nonlinear SSMs using Gaussian process optimisation}}
\small
\textsc{Inputs:} Algorithm \ref{alg:APFlike}, $K$ (no.\ iterations) and $\theta_1$ (initial parameter). \\
\textsc{Output:} $\widehat{\theta}$ (est.\ of the parameter).
\algrule[.4pt]
\begin{algorithmic}[1]
	\STATE Initialise the parameter estimate in $\theta_1$.
	\FOR{$k=1$ to $K$}
		\STATE Sample $\widehat{\ell}(\theta_k)$ using Algorithm \ref{alg:APFlike}.
		\STATE Compute \eqref{eq:GPpost} and \eqref{eq:GPpred} to obtain $\ell(\theta)|\mathcal{D}_k$.
		\STATE Compute \eqref{eq:mumax} to obtain $\mu_{\max}$.
		\STATE Compute \eqref{eq:EIdef} to obtain $\theta_{k+1}$.
	\ENDFOR
	\STATE Compute the maximiser $\mu(\theta|\mathcal{D}_K)$ to obtain  $\widehat{\theta}$.
\end{algorithmic}
\label{alg:FullAlg}
\end{algorithm}

\section{Numerical illustrations}
\label{sec:results}
Finally, we are ready to combine the methods discussed in the previous three sections into the final algorithm and it is presented in Algorithm \ref{alg:FullAlg}. In the following, we use an LGSS model and a nonlinear model to illustrate the behaviour and the performance of the proposed algorithm. We compare the proposed method in the latter model with the SPSA algorithm \cite{Spall1987}. This algorithm is selected as it also only makes use of zero-order information (the log-likelihood estimates) and is known to perform well in many problems, see e.g.\ \cite{Spall1998}.

\subsection{Implementation details}
For the GP, we use a constant mean function and the Mat\'{e}rn kernel with $\nu = 3/2$. Note that, other choices of mean functions and kernels (especially the combination of kernels) can possibly improve the performance of the algorithm. This is especially important in models where the log-likelihood in non-isotropic.

The \textit{GPML toolbox} \cite{RasmussenWilliams2006} is used for estimation of the hyperparameters by EB and for the computation of the predictive distribution in \eqref{eq:GPpost}. For the acquisition function, we use the EI with $\zeta=0.01$ following the recommendations in \cite{Lizotte2008}.

The optimisation in \eqref{eq:EIdef} is non-convex and therefore difficult to carry out in a global setting. Two common approaches in GPO are to use multiple local search algorithms in a Monte Carlo setting \cite{Lizotte2008} or using a global optimisation algorithm \cite{BrochuCoraDeFreitas2010}. In this paper, we use the latter method with the gradient-free DIRECT global optimisation algorithm \cite{Jones1993} and the implementation written by Daniel E.\ Finkel, available from \url{http://www4.ncsu.edu/~ctk/Finkel_Direct/}. A maximum of $500$ iterations and (cheap) evaluations of the surrogate function are used in the DIRECT algorithm for each optimisation.

\subsection{Linear Gaussian state space model}
\label{sec:resLGSSM}
\begin{figure}[!t]
	\centering
	\includegraphics[width=\columnwidth]{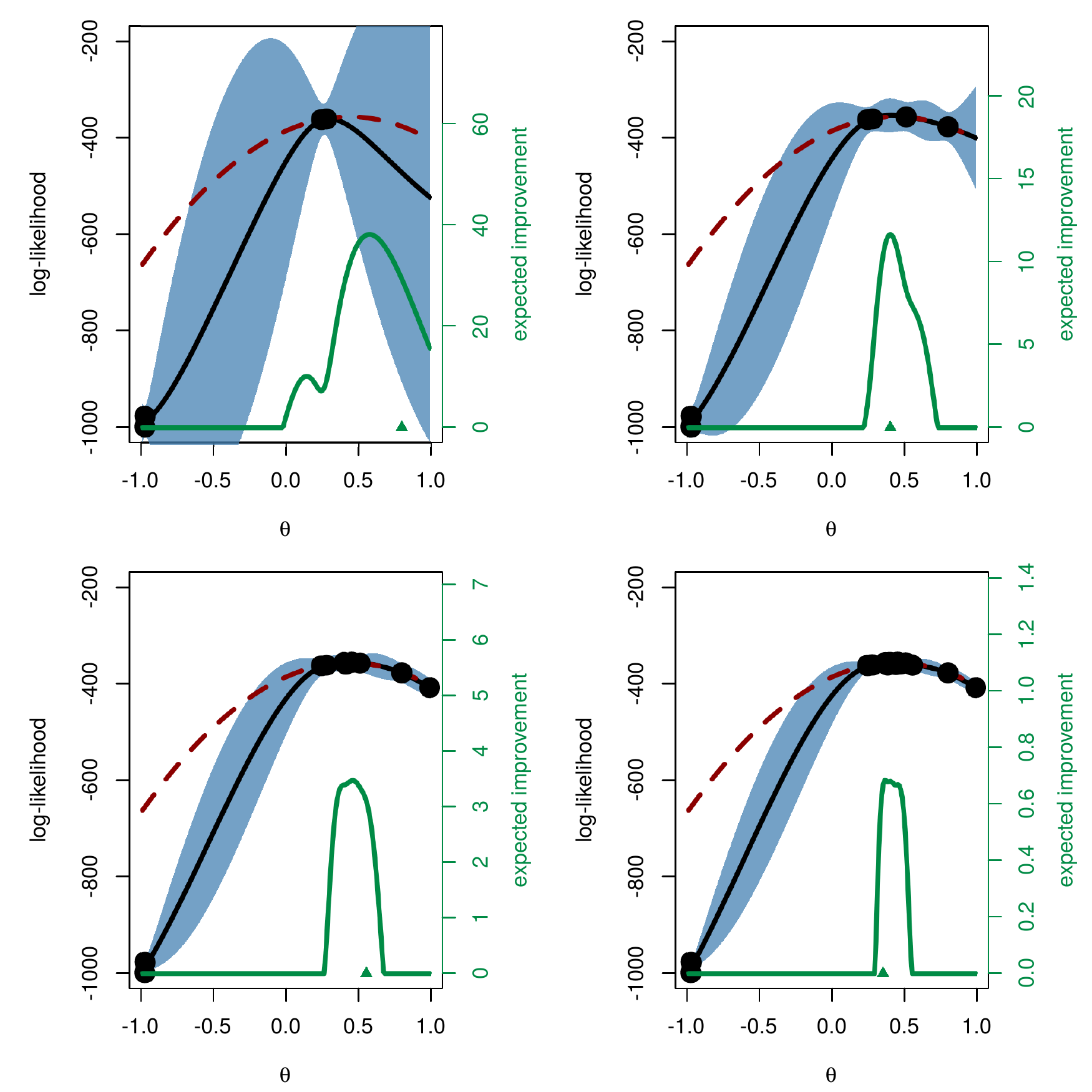} 
	\caption{\small{The surrogate model (solid line) and EI (green line) at iterations $\{5,10,15,50\}$ for the LGSS model. The true log-likelihood is presented as the dashed red line. The $95\%$ confidence of the surrogate function is marked by blue. "$\bullet$" and "$\triangle$" indicate samples from the log-likelihood and the maximum of the EI obtained by the DIRECT alg.\ }}
	\label{fig:resultsLGSSM}
\end{figure}

We begin with the LGSS model using one parameter in \eqref{eq:LGSSMmodel}, as this enables us to investigate the behaviour of the proposed algorithm in detail. We use $N=1 \thinspace 000$ particles, $K = 50$ iterations and the initial parameter $\theta_1=-0.98$. In Figure \ref{fig:resultsLGSSM}, we present the surrogate function and the expected improvement at different iterations. The algorithm converges rather quickly for this simple toy example with the parameter estimate $\widehat{\theta}=0.48$. As a comparison, the MLE obtained by the Kalman filter by maximisation on a grid of parameter values is $\theta_{\text{MLE}}=0.44$.

\subsection{Nonlinear stochastic volatility model}
Consider the Hull-White stochastic volatility model \cite{HullWhite1987},
\begin{subequations}
\begin{align}
	x_{t+1}|x_t  &\sim  \mathcal{N} \left( x_{t+1}; \theta_1 x_t , \theta_2^2 \right), \\
	y_{t}|x_t    &\sim  \mathcal{N} \left( y_t; 0, 0.7^2 \exp(x_t) \right),
\end{align}%
\label{eq:CIRSVmodel}%
\end{subequations}%
\noindent where the parameters are $\theta^{\star}=\{\theta^{\star}_1,\theta^{\star}_2\}=\{0.90,0.20\}$. We use $\Theta=\Theta_1 \times \Theta_2 = [-1,1] \times [0,2]$, $T=250$ time steps, $N=1 \thinspace 000$ particles, $K = 300$ iterations and the initial parameter $\theta_1=\{0.5,0.5\}$. We implement the SPSA algorithm as suggested by \cite{Spall1998} using the recommended settings for the parameters $\alpha$, $\gamma$ and $C$. We manually tune the parameters $a=0.03$ and $c=0.04$ to achieve good performance for our problem.

The GPO algorithm again converges rather quickly after about $50$ evaluations of the log-likelihood and returns the parameter estimate $\widehat{\theta}=\{0.896,0.187\}$. The SPSA algorithm converges slower and requires more than $200$ evaluations of the log-likelihood to reach the neighbourhood of the true parameters. Even more iterations are required for the estimates to stabilise. This shows, for this particular example, that the GPO algorithm could be a competitive choice for maximum likelihood estimation.

\begin{figure}[ht]
	\centering
	\includegraphics[width=\columnwidth]{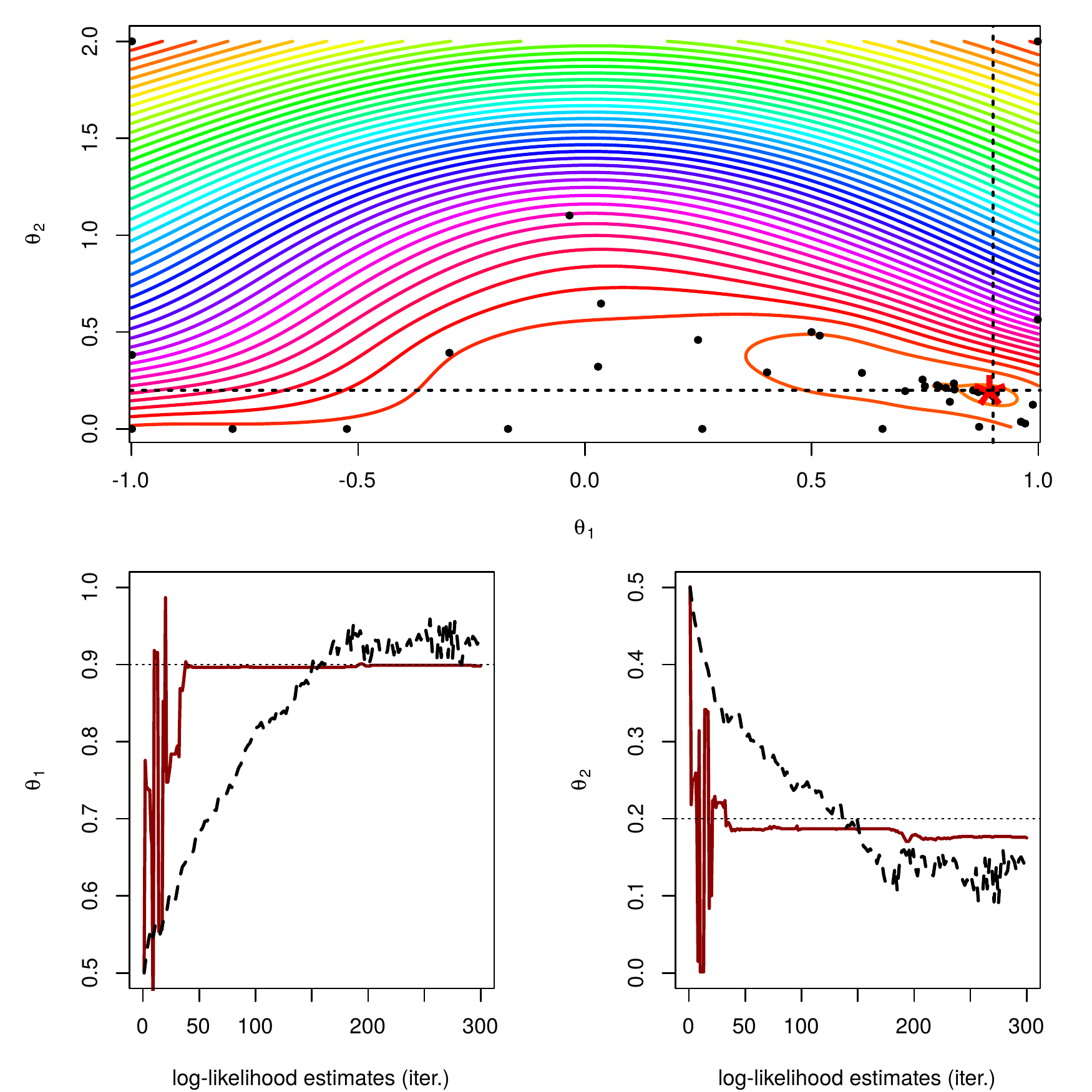} 
	\caption{\small{Upper: the log-likelihood model generated using the $K$ iterates with the est.\ parameters (red star). Lower: the estimates of $\theta_1$ (left) and $\theta_2$ (right) using GPO (solid) and SPSA (dashed). The true parameters are presented by dotted lines.}}
	\label{fig:resultsHWSV}
\end{figure}

\section{Conclusions}
The results in the previous section indicate that the proposed method does not require many estimates of the intractable log-likelihood. This is due to the GP model that captures the overall structure well and enables an efficient sampling mechanism in the form of the acquisition rule. With this and the comparison with SPSA in mind, we hope that this algorithm shall turn out to be a competitive alternative to more advanced algorithms.

Important future work includes benchmarking of the proposed method, alternative acquisition rules and investigating possibilities for bias-compensation of the log-likelihood estimate. Also, the Gaussian process models can be useful as an alternative to compute the gradient (score function) and negative Hessian (the observed information matrix) of the log-likelihood. Estimating the latter is an important problem in e.g.\ nonlinear input design, and this approach could decrease the variance in such estimates.

At \url{http://users.isy.liu.se/en/rt/johda87/}, we provide source code to reproduce some of the numerical illustrations in this paper.

\section*{Acknowledgements}
The authors would like to thank Prof.\ Thomas B.\ Sch\"{o}n, Dr.\ Carl E.\ Rasmussen, Roger Frigola and Andrew McHutchon for interesting discussions and suggestions that greatly improved this paper. 

\clearpage
\bibliographystyle{plain}
\bibliography{dahlin}
\end{document}